# A flexible solution to embrace Ranking and Skyline queries approaches

Simone Censuales

Politecnico di Milano
Milan, Italy
simone.censuales@polimi.it

March 5, 2022

## Abstract

The *multi-objective optimization problem* has always been the main objective of the principal traditional approaches, such as Ranking queries and Skyline queries. The conventional idea was to either use one or the other, trying to exploit both ranking queries advantages when it comes to taking into account user preferences, and skyline queries points of strength when the main objective was to obtain interesting results from a dataset in a simple, yet effective fashion, both of them showing limitations when entering specific fields of interest. Even though some interesting new techniques were proposed to solve those problems, this paper's main aim is to explore the world of *Flexible/Restricted Skylines*, a way to combine the concept of dominance with a set of scoring functions of interest, pointing out the effectiveness of new operators involved and underlining the main differences in performance with the traditional approaches.

## 1 Introduction

Coping with the problem of optimizing different criteria in a simultaneous fashion has always been a central topic when approaching all data-intensive tasks. This kind of problem is known as *multi-objective optimization*. As stated in [1] and [3], in order to deal with this issue, three main approaches are usually taken into account:

- The *lexicographic approach*[15], where a strict priority is provided out of all the possible attributes
- The *ranking* (or *top-k*) *queries approach*[16], in which a *scoring function* S is exploited to assign a numerical ranking score to each tuple involved in the process. In this way, the previous multi-objective problem is restricted to a single-objective one
- The *skyline approach* [10], which returns a set of potentially valuable objects according to the concept of *dominance*.

Despite their usefulness, all these methods have shown some limitations in some specific contexts, analyzed in further sections of the survey.

In this article, we provide a short, yet consistent, overview of the two latter approaches, taking into account the pros and the cons of each proposal. Furthermore, an interesting, but brief overview of some innovative attempts to best the principal issues are indeed provided. Finally, the core of the paper is represented by the notion of "*Flexible/Restricted Skyline*" and its applicability in solving both the multi-objective optimization problem and overcoming the problems stated for previous paradigms. Table 1 shows the accomplishment of some major evaluation criteria by each approach.

To summarize, the main contributions of the survey are:

    (i)   We introduce the multi-objective problem and the major paradigms to address it.

    (ii)  We consider the Ranking queries approach, underlining its points of strength and limitations.

    (iii) We examine the Skyline queries approach, showing its benefits and drawbacks.

    (iv) We analyze the major aspects of Flexible/Restricted Skylines, also providing a general view of the algorithms used to implement it.

    (v)  We provide a critical comparison of the potential advantages of Flexible/Restricted Skylines over Ranking queries and traditional Skyline queries.

    (vi) We introduce a list of convenient alternative paradigms, which further extend the already presented approaches.

|  | RANKING (TOP-K) QUERIES | SKYLINE QUERIES |
|---|---|---|
| *SIMPLICITY* | ✗ | ✓ |
| *INTERESTING RESULTS* | ✗ | ✓ |
| *CARDINALITY CONTROL* | ✓ | ✗ |
| *ATTRIBUTES TRADE-OFF* | ✓ | ✗ |
| *IMPORTANCE OF ATTRIBUTES* | ✓ | ✗ |

*Table 1: Evaluation Criteria*

## 2 Ranking (Top-*k*) Queries

Ranking (or top-k) queries are the most common approach when dealing with multi-criteria decision making. Its principal aim is to retrieve only the *k* best options from a (possibly) large result set, according to a <u>user-defined scoring function</u>, such as the *weighted summation of multiple scores*. In this way, it is possible to integrate users' preferences in terms of parameters of the function.

Given a multi-dimensional dataset, only the *k*-best tuples become part of the result, but if more than one set of *k* tuples satisfies the ordering, any of those is a valid answer, and this is what is commonly addressed as *non-deterministic semantics*.

For a better understanding, let's consider the following example:

Alice wants to buy a car with high Horsepower (HP) and with high Miles Per Gallon (MPS) and, so, she visits a large car database. Since it would be quite challenging to go through all the car tuples in the database to find a set of potentially good cars, Alice decides to use the utility function (scoring function assumed to be already provided), to address her preferences. For each tuple of the database, the utility is computed and only those *k* tuples with the highest utilities are returned. In Table 2 we can observe a meaningful representation ($w_i$ represents the weight value and $V_i$ represents the database value).

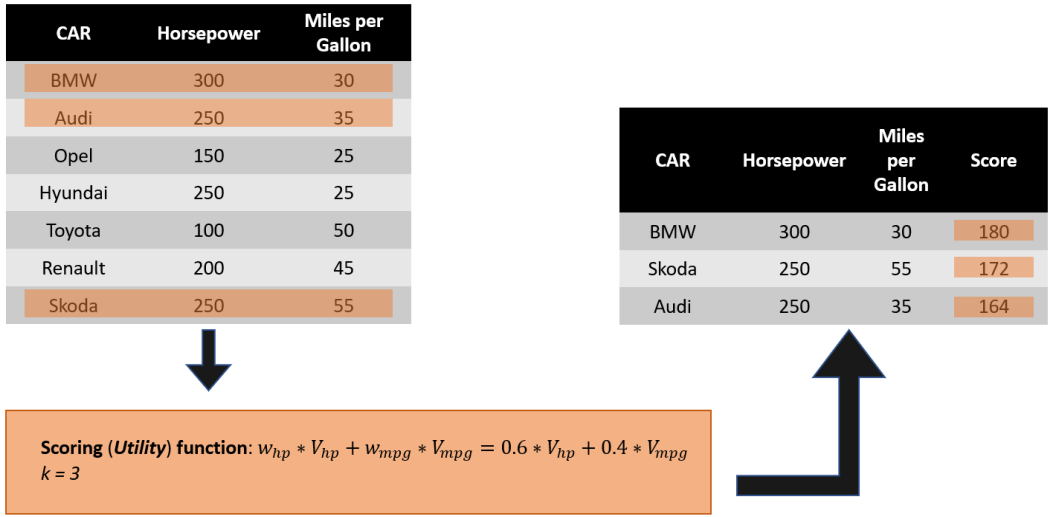

*Table 2: Ranking Query*

The main advantages of this solution are the *control of the cardinality*, which allows the user to customize the number of tuples given in the output of the query, and the possibility to establish a *trade-off* among attributes, so that users can properly address their preferences. Anyway, as pointed out in [1], the result set of a ranking query depends on the way user preferences are converted into the parameters of the scoring function. Furthermore, it is hard to foresee the impact on ranking of adjusting one or more parameters.

# 3 Skyline Queries

In order to properly describe the Skyline queries approach, it is necessary to introduce the concept of *dominance* [1, 2, 3, 8, 10]: a tuple *t* dominates another tuple *t'* if:
- *t* is not worse than *t'* on any of the attributes
- *t* is strictly better than *t'* on at least one of the attributes.

Formally, a skyline set [10] contains all objects that are not dominated by any other object in the dataset. They are useful because they can provide users with potentially interesting results without the need of specifying a customized weight for each criterion.

For a better understanding, let's refer to the previous car example:
Car *c* (BMW = HP: 300 and MPG: 30) dominates car *c'* (Hyundai = HP: 250 and MPG: 25) since its utility is higher than the second car, no matter what function Alice decides to use. We can observe a meaningful representation in Table 3.

The main benefits of this approach are the *effectiveness* in identifying interesting objects, even if users' preferences are not known, and its *simplicity*. On the other hand, it does not allow users to control the size of the output and it does not consider what really matters for the user.

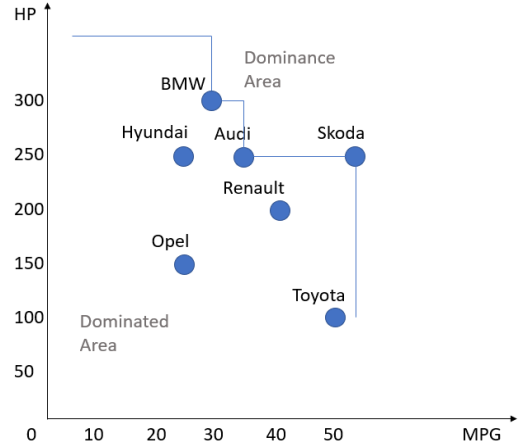

Table 3: Skyline Query

# 4 Flexible/Restricted Skylines

All the analyzed approaches have shown some major limitations, which could interfere with an optimal user experience. The main objective of the paper is to present a new improved paradigm, called *Flexible* (or *Restricted*) *Skyline*, which aims at overcoming those issues, also combining the point of strength of *Ranking queries* and *Traditional Skyline queries*.
In order to properly describe the solution, we provide a generalization of the multi-objective optimization problem:

**Problem 1**: *Given N objects described by d-dimensional attributes in a dataset and given some notion of "quality of data", possibly conflicting, for each attribute, find the best objects.*

We have seen so far how the most important paradigms approach the Problem 1, in particular:
- *Ranking-like approaches*: the output must consider the relevance of attributes
- *Skyline-like approaches*: the output must consider dominance among attributes.

Restricted skylines approach proposes a flexible framework [1, 2, 3] able to embrace user preferences through the employment of constraints.
In particular, its behavior concerns the application of the dominance paradigm to a limited set of monotone scoring functions. In doing so, it extends the notion of dominance introduced in traditional Skyline queries to also involve the considered set of functions, by means of *F-Dominance*:

F-DOMINANCE [3]: *let F be a set of monotone scoring functions. A tuple t F-dominates another tuple s ≠ t, if and only if for every function f ϵ F → f(t) ≤ f(s), denoted by $t \prec_F s$.*

For clarity purposes, an example is provided:
Let's consider the two different tuples t = ⟨1, 5⟩ (*green dot in the referenced Figure 4*) and k = ⟨2, 4⟩ (*blue dot in the referenced Figures*), the monotone scoring functions f1(x, y) = $x^2 + y$ and f2(x, y) = $x + y$ in the following intervals:
- $x \rightarrow [0, +\infty)$
- $y \rightarrow [0, +\infty)$

as well as the set F = {f1, f2} and the set M = {f1, f2, f3}. We have $t \prec_F k$, since f1(t)=6 < f1(k)=8 (in Figure 4) and f2(t) = f2(k) = 6 (in Figure 5), and therefore the condition of F-Dominance definition holds.

However, t $\not\prec_M$ k, since M includes, among others, f3(x, y) = $x + y^2$, for which f3(t) = 26 > f3(k) = 18 (in Figure 6), and so it violates the condition of F-Dominance definition.

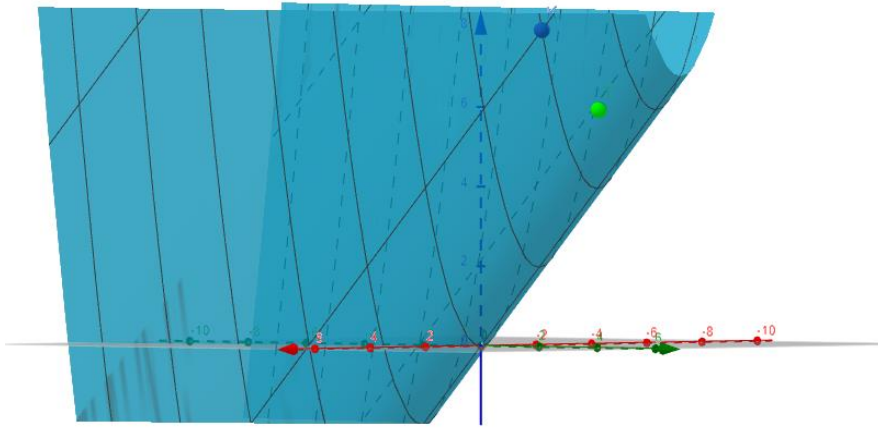

*Figure 4: f-dominance f1(x,y)=x^2+y*

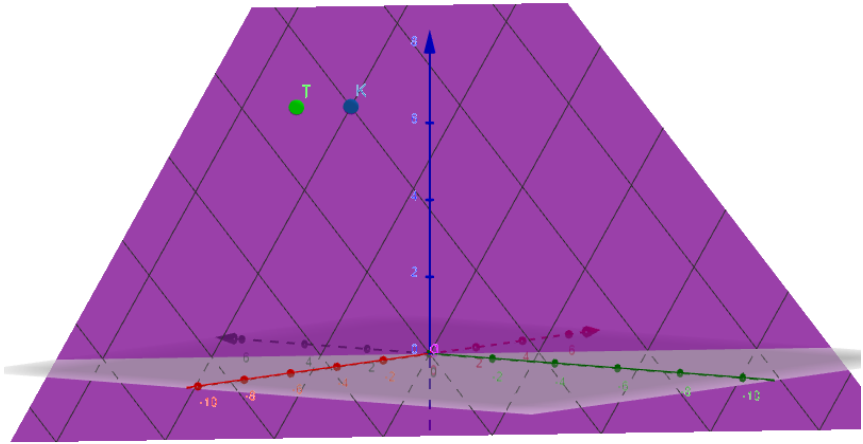

*Figure 5: f-dominance f2(x,y)=x+y*

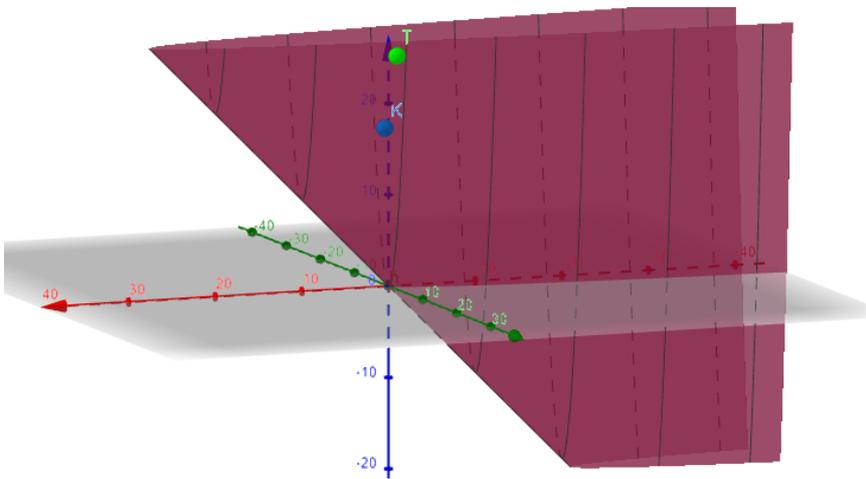

*Figure 6: f-dominance f3(x,y)=x+y^2*

Bearing in mind the *F-Dominance* definition, we can indeed introduce two F-Skyline operators, called *non-dominated restricted skyline* (ND) and *potentially optimal restricted skyline* (PO) respectively.

The ND operator consists of the set of tuples which are not F-dominated in an instance *r* of the considered relation schema *R*. More formally, we can state:

$$ND(r; F) = \{t \in r \mid \nexists\, s \in r, s \prec_M t\}$$

The PO operator gives back the set of tuples considered as *best tuples* according to one or more scoring function in F. More formally, we can affirm:

$$PO(r, F) = \{t \in r \mid \exists\, f \in F, \forall\, s \in r, s \neq t \rightarrow f(t) < f(s)\}$$

Going back to the previous car example, we can identify ND and PO outputs. In particular, according to their respective definitions, we have to point an instance *r* of the relation schema to determine the set of tuples found by ND operator, and (at least) a scoring function to retrieve the set of tuples spotted by PO operator:
- the instance *r* can concern only the Horsepower and MilesPerGallon attributes of the overall relation schema (so as to match our previous graph), which would realistically contain further attributes
- the scoring function $f(x_1, x_2) = 0.8 * x_1 + 1.6 * x_2$ will be considered to compute PO set of tuples, where $x_1$ is the value of Horsepower and $x_2$ is the value of MilesPerGallon.

We can observe the visual representation in Figure 7.

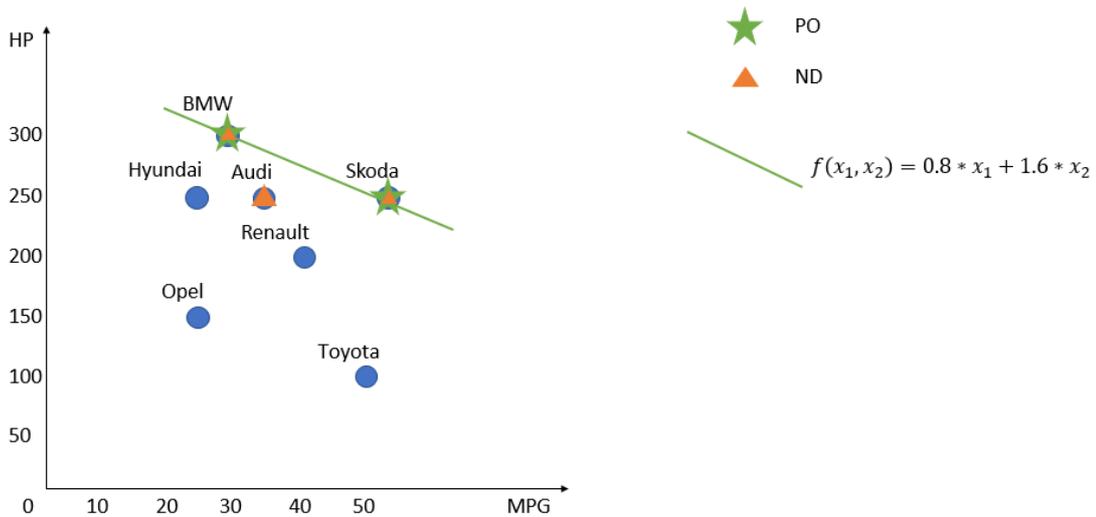

*Figure 7: ND and PO operators representation*

An interesting comparison can be made between the two operators previously introduce and the traditional skyline approach; with this intent, let briefly introduce the *dominance* notion in traditional skyline queries:

DOMINANCE IN SKYLINE [1]: *let s and t be tuples (described as a set of attributes $A_i$) over a relational schema R. Being r an instance of R, we can say that t dominates s (written as $t \prec s$) if both of the following statements are verified:*
- $\forall\, i, 1 \leq i \leq d \rightarrow t[A_i] \leq t[A_i]$
- $\exists\, j, 1 \leq j \leq d$ and $t[A_j] < s[A_j]$

Indeed, the skyline of r (called SKY(r)) is defined as: $SKY(r) = \{t \in r \mid \nexists s \in r \rightarrow s \prec t\}$.

As an immediate aftereffect to previous definitions, we can assert [3] that considering any set F of *monotone* scoring function, the set of tuples identified by both ND and PO are included in the set returned by SKY, formally:

$$PO(r; F) \subseteq ND(r; F) \subseteq SKY(r)$$

This is an important statement because it clearly proves that the size of the sets identified by ND and PO is:
- *Smaller* with respect to the set of tuples returned by SKY in the general case
- *Equal* to SKY tuple set considering the worst case.

## 4.1 Algorithms for computing operators

### 4.1.1 Computing ND operator

We can identify some procedure steps to compute ND operator:
1. *Sorting Decision Step*: at first, we need to decide whether to sort or not the dataset, so as to produce a topological sort [3] with respect to the F-dominance relation. If an algorithm counts on sorting, it is indicated by the letter "**S**" in its name, otherwise by the letter "**U**".
2. *F-Dominance Step*: in order to test the F-dominance relation between tuples we can consider two different approaches:
    (i) Solve a minimizing Linear Programming problem concerning a set *C* of linear constraints, indicated by "**LP**" in the algorithm name
    (ii) Using vertex enumeration to check whether a tuple *t* belongs to the F-Dominance region of tuple *s*, indicated by "**VE**" in the algorithm name
3. *Phases Step*: in this step, we should decide if the computation of ND takes place after computing the SKY operator (in this case the algorithm name will have a "**2**" in the name) or if the dominance and F-dominance tests need to be integrated (identified with a "**1**" in the algorithm name).

Let's consider four of the main algorithms used to compute ND operator, according to different approaches: SVE1 - SVE2 and ULP1 - ULP2. As claimed in [3], sorted strategies are proven to be better than unsorted one, especially when dealing with large datasets, whereas, as far as LP against VE comparison is concerned, vertex enumeration is faster by far than linear programming.

### 4.1.2 Computing PO operator

The most beneficial way to compute PO is [2] starting from the tuples found in ND and discarding those that are F-dominated by any convex combination of other tuples in ND. An optimal way to directly check the F-domination is to reduce the potentially optimal set of tuples as soon as possible, through the following heuristic:
1. First, let's consider a convex combination of just two tuples and test for F-dominance between them
2. After each round, double the number of tuples taken into account
    2.1 As long as the number of tuples does not overcome the number of |PO – 1|, we can prune the tuples from the considered PO set (*sufficient condition*)
3. Enumerate each candidate tuples proven to be F-dominated in PO in reverse order, as they are most likely to be F-dominated
4. Check the existence of a convex combination of the first tuples in PO by using a linear system. Those tuples have been proven to be the most likely to F-dominate the others
5. Prune every tuple which is proven to be F-dominated by any convex combination of the remaining tuples.

The algorithm already described is called POND (computing PO via ND) and it has been proved to be fairly effective [2], especially for the sufficient condition for pruning tuples in the first steps of the. Furthermore, POND algorithm has been proven to be faster in terms of time complexity even against LP approaches.

# 5 Comparison

## 5.1 F-Skylines versus Ranking queries

In order to compare properly the two already analysed approaches, we will refer to two measures:
- *Precision (PRE)*: given a set $T$ of top-k tuples in relation $r$ with respect to a scoring function $f$ [1], its precision regarding a set $S$ is defined as: $PRE(S) = |S \cap T_k(r; f)| / k$
- *Recall (REC)*: given a set $T$ of top-k tuples in relation $r$ with respect to a scoring function $f$, the recall is defined as: $REC(S) = |S \cap T_k(r; f)| / |S|$

In the context where the set $S \in \{SKY, ND, PO\}$ previously analysed, [1] claims that a high precision would mean that the majority of the top-k tuples belongs also to $S$, whereas a high recall would indicate that the majority of tuples in S are also top-k query.

When considering a scoring function $f$ which represents a *weighted sum*, we can choose the centroid of the polytope $W(C)$, where $C$ is any set of linear constraint on weights, as its related weighted vector. This will allow us to consider both $f \in F$. For every value of $k$ considered, the ranking queries will retrieve non-$F$-dominated, also potentially optimal, tuples more easily than F-skylines[1]:

$$PRE(PO) \leq PRE(ND) \leq PRE(SKY)$$

For recall measure considerations, the results are very similar to those provided by the analysis of precision.
In conclusion, we can convey that F-Skylines main advantage is the ability to provide interesting results to the user, in exchange for a higher computational overhead compared to Ranking queries, which, in most of the [1] tests only require a small fraction of F-Skylines' algorithm execution time.

## 5.2 F-Skylines versus Traditional Skyline queries

According to dataset analysis in [3], F-Skylines outperforms traditional Skyline queries in every scenario (underlining that PO operator has proved to be more effective than ND), especially when more constraints are considered.
Regarding execution time, the computation of SKY requires, in most cases, the same time as computing ND, with a small advantage for the latter when considering 1-phase algorithms.
We can indeed conclude that F-Skylines do not introduce any expensive overhead, while being very successful in reducing the size of the output, with respect to traditional Skyline queries.

# 6 Related Works

Since neither the top-$k$ queries nor skyline queries can flawlessly address the multi-objective optimization problem, a large number of new techniques were proposed by the scientific community to either solve those issues proper of the previous approaches or to introduce interesting new paradigms.

## 6.1 Attempts to overcome Ranking queries limitations

**Top-$k$ queries with uncertain function** [12]**:**

In many situations, choosing an uncertain preference specification language might be a better solution to address users' preferences.
The uncertainty could be beneficial even to those situations where users are able to provide the right weights to the scoring function since it would make it possible to analyze the sensitivity of the ordering with respect to changes in the chosen weights.
When a user tries to calibrate the scoring function, he/she could be more interested in assessing the biggest acceptable change in the chosen weights that do not affect the computation, instead of perfectly nailing the correct static weights.

Anyway, this approach still presents basically the same limitation [14] as ranking queries, as it is dependent upon a set of measures that could not perfectly match users' preferences.

**Uncertain Top-*k* query** [4]**:**
This approach is very similar to the previous one, as it takes also into account the uncertain weight concept, but it extends it even further.
We can distinguish three different inputs for UTK (*Uncertain Top-K query*):
1. A dataset *D*, containing a set of records/tuples *p*
2. A positive integer *k*
3. A region R of interest, located inside the preference domain, where each tuple can be considered as a vector

When $k = 1$, the score of each record *p* is calculated thanks to a given input vector *w* of *weights*, assuming that higher values are preferable for every attributes.
In this way, we are able to distinguish two main derivations, respectively called UTK1 and UTK2:
- UTK1 directly addresses the set of minimal records that may rank among the top-*k* tuples, when the vector containing all the weights lies inside the region R
- UTK2 gives in output the correct top-*k* set of tuples, for each weight vector inside the region R

The main disadvantage of the paradigm is that processing time can be slowed when the computation includes a large set of attributes.

**Regret minimization** [8, 13]**:**
This paradigm allows the user to control the output size, without requiring any utility function. Its main point of view concerns the fact that when we try to restrict the output size, there is a chance of having a difference between the highest utility value among all tuples in the database and the highest utility value among all the chosen *k* tuples.
The ratio representing this gap is called *regret ratio*, a real number between 0 and 1. In particular, the closer to 0 the ratio is, the happier the user is.
The starting point to measure such regret ratio is to consider a way to capture users' happiness. We are able to do it with the so-called *utility function* $f(x)$, where given a point *p* represents a possibility inside the dataset and $f(p)$ is the correlated utility. It is not possible to know the utility function used by the user to give a numeric value of his/her preference to all the tuples, also because, most of the times, even the user itself is not aware of it. The only fact that we are able to state is that every user tries at maximizing his/her respective utility. Under this statement, we can introduce the notion of *gain*, used to identify user's utility when considering only a subset of tuples. So, given a subset S of the original database D, the gain is defined as those tuples *p* capable of achieving the maximum of utility, according to the *utility function f*, or, more formally;

$$gain(S, f) = \max_{p \epsilon S} f(p)$$

Furthermore, we call the *maximum regret ratio* the biggest possible ratio among all users, and it represents how regretful a user is when getting the best tuple among all the selected *k*, but not the best tuple among all those in the database.
The main objective of a *k*-regret query is to identify and get a set of *k* tuples, so as to minimize the maximum regret ratio of the set itself.
Its main limitations are that most of the algorithms employed only deal with linear utility functions, without considering monotone or generic ones, and even further, they require the interaction with the user to properly work.

## 6.2 Attempts to overcome Skyline queries limitations
**Skyline ordering** [11]**:**
Although this paradigm takes into account also some rationale from top-*k* queries, we consider it more of a way to overcome the main shortcomings of skyline queries.

This approach introduces a partitioning of a given dataset. The partitions will be further ordered, reserving the possibility for the use of set-wide maximization techniques, in which a portion of the skyline is selected in a way that it is possible to maximize a collective quantity-based objective, among those partitions. Doing so, it allows supporting arbitrary size constraints over the traditional skyline query.

The opening of the procedure is to start from the first partition and then continue to output further partitions until at least *k* of those have been issued (top-*k* logic). The last partitions found are then pruned so that exactly *k* points can be returned. This guarantees flexibility in the computation over the queries.

The main limitation of this framework is the lack of local optimization within each partition.

**ε -Skyline [7]:**

This approach is born to overcome three main limitations of the traditional skyline paradigm:
- Lack of cardinality control
- Skyline tuples are not comparable among them, and they lack a semantic order
- Invariant to dimensional scaling, since the dominance of a tuple with respect to another cannot be altered by reducing or increasing the weight of dimension.

The ε-Skyline defines a *flexible* dominance relation, which provides a better semantics and a better way to control skyline tuples. Moreover, the monotonic function used in this solution varies from the empty set to the whole dataset, allowing the user to alter it by changing the values of the weights employed in the computation.

The main limitations of this approach are the partial customizability of the utility function and, in some relevant cases, the complexity required can be greater than other techniques' computability.

## 6.3 Attempts to combine the approaches or to propose new solutions

**ORD-ORU [5]:**

Similar to what ε-Skyline already introduced, ORD and ORU operators' main objective is to uphold three hard requirements that limit the experience of both ranking queries and skyline ones:
- Lack of cardinality control, concerning the impossibility for skyline to specify the size of the output
- Personalization of the proposed solution, regarding the lack of consideration of user specific preferences in traditional skyline approach
- Relaxed preference input, which should take into account flexibility in the stated preferences.

Both of them expand the preference input, besides accomplishing a good level of responsiveness and scalability.

ORD operator sticks closer to skyline paradigm, whilst ORU employs more of a ranking-based approach.

The main issues with the two operators are that both do not give a ranked output and they only offer partial customizability. Furthermore, even if ORD can be employed for those applications that require sub-second responses, ORU is not so ready, because it would require some sort of parallelization.

Even though they are not as simple to implement as F-Skylines, they are able to offer better overall performance, especially when considering multi-dimensional computation on sparse dataset [1] [5].

**Pairwise Comparison [9]:**

This innovative approach bases its main objective on the "*Thurstone's Law of Comparative Judgement*", which states that a pairwise comparison is a more effective way to learn a preference function, with respect to either choosing a set of preferred entities or determining the overall ranking.

*Rational scheme*: we select a set of entity-attribute pairs and ask the user which of those are preferable among those. We then estimate the final preference function according to the feedback.

For such a preference scheme to be addressed, it must encounter three conditions:
- Easy to answer
- Only few comparisons should be needed in order to learn an approximation of the preference function
- System should be fast at posing the comparisons

The main drawback of this solution is that the user still needs to perform lots of comparisons so as to gain reasonable precision, all resulting in a non-negligible computation complexity.

In Figure 8 it is provided an overall comparison of all the approaches analized so far, regarding the main dimension of comparison.

| APPROACH \ DIMENSIONS | Cardinality Control | Interesting Result | Simplicity | Attributes Trade-Off | Performance Efficiency |
|---|---|---|---|---|---|
| Ranking Queries | ✓ | | | ✓ | ✓ |
| Skyline Queries | | ✓ | ✓ | | |
| F-Skylines | (✓) | ✓ | ✓ | ✓ | |
| Pairwise Comparison | | ✓ | ✓ | ✓ | |
| Uncertain Top-$k$ Queries | | ✓ | ✓ | ✓ | |
| Regret Minimization | ✓ | | ✓ | ✓ | |
| Skyline Ordering | (✓) | ✓ | ✓ | | |
| ε-Skyline | ✓ | ✓ | | ✓ | |
| ORD-ORU | ✓ | ✓ | | ✓ | |

Figure 8: Table of comparison of all approaches

# 7 Conclusion

In this survey, we have presented the *multi-objective optimization problem*, as well as the classical approaches to deal with it: Ranking queries and Skyline queries, underlining both advantages and drawbacks. Then, we went through the analysis of interesting proposals to overcome the main issues found for traditional approaches. More specifically, among all the proposals, we identified those techniques which address the Ranking queries flaws, those which try to solve Skyline queries weaknesses and, in the end, those whose main objective is to either combine the two approaches in an innovative way, or to propose a new view of the problem, always pointing at their respective pros and cons.

Furthermore, we have introduced the notion of *Flexible/Restricted Skyline*, which main aim is to marry the skyline paradigm with the capability of Ranking queries to address user preferences, using constraints on the weights of scoring function. For this purpose, we introduced the notion of F-Dominance regarding a family $F$ of (possibly monotone) scoring functions. Moving on, we addressed the two main operators of Flexible Skylines, through the notion of *non-dominated* (ND) and *potentially-optimal* (PO) tuples with respect to $F$.

To propose an initial comparison with previous techniques, we briefly introduce Skyline operator SKY and we assessed some relations among all the operators described. In section 5.1 we addressed the most relevant algorithms to compute Flexible Skylines operators and their relative field of applicability, underlining the advantages of one approach with respect to another.

In conclusion, a comparison with the traditional techniques has been offered, so that it would be possible for the reader to better understand the reason why Flexible/Restricted Skylines have been proposed as a new optimal solution to the original multi-objective optimization problem.